\documentclass[pra,aps,twocolumn,byrevtex,showpacs,tightenlines,superscriptaddress]{revtex4}
\usepackage{epsfig}
\begin{document}

\title{Qudit Quantum State Tomography}
\author{R.~T.~Thew}\email{Robert.Thew@physics.unige.ch}

\affiliation{Center for Quantum Computer Technology, University of
Queensland, QLD 4072, Brisbane, Australia}

\affiliation{Group of Applied Physics,University of Geneva,20 rue
de l'Ecole-de-M\'edecine,CH-1211 Geneva 4,Switzerland }

\author{K.~Nemoto}%\email{nemoto@informatics.bangor.ac.uk}
\affiliation{School of Informatics, University of Wales, Bangor,
LL57 1UT, United Kingdom}

\author{A.~G.~White} %\email{andrew@physics.uq.edu.au}
\affiliation{Center for Quantum Computer Technology, University of Queensland, QLD 4072, Brisbane, Australia}

\author{W.~J.~Munro} %\email{bill_munro@hp.com}
\affiliation{Hewlett-Packard Laboratories, Filton Road, Stoke
 Gifford, Bristol BS34 8QZ, United Kingdom}

\date{\today}

\begin{abstract}
Recently quantum tomography has been proposed as a fundamental
tool for prototyping a few qubit quantum device. It allows the
complete reconstruction of the state produced from a given input
into the device. From this reconstructed density matrix, relevant
quantum information quantities such as the degree of entanglement
and entropy can be calculated.  Generally orthogonal measurements
have been discussed for this tomographic reconstruction. In this
paper, we extend the tomographic reconstruction technique to two
new regimes. First we show how non-orthogonal measurement allow
the reconstruction of the state of the system provided the
measurements span the Hilbert space. We then detail how quantum
state tomography can be performed for multi qudits with a specific
example illustrating how to achieve this in one and two qutrit
systems.
\end{abstract}
\maketitle

\section{Introduction}
With increasing interest in quantum computing, cryptography, and
communication, it is of paramount importance that there exist
means of benchmarking quantum information experiments.  A
singularly useful tool in this regard is Quantum State Tomography
(QST), which provides a means of fully reconstructing the density
matrix for a state.  The procedure relies on the ability to
reproduce a large number of identical states and perform a series
of measurements on complimentary aspects of the state within an
ensemble. The concept is not new, with the first such techniques
developed by Stokes \cite{Stokes52a} to determine the polarization
state of a light beam.  Recently James {\it et al.}
\cite{James01a} gave an extensive analysis of qubit systems
specifically focusing on polarization entangled qubits, building
on earlier experimental work \cite{White99a}, but more generally
for any number of qubits.  We also refer the reader to Leonhardt`s
book \cite{Leonhardt97a} which gives an introduction to some of
the concepts and experimental techniques of tomography relating to
continuous variable systems in modern quantum optics.

It is our aim here to expand on the work of James {\it et al.} in
two ways: firstly, to detail how to perform QST on systems of $n$
qudits; secondly, to show how to perform QST when access to a full
range of single qubit rotations and hence the state space is
restricted.  The first point is also motivated with respect to
fundamental questions regarding non-locality in higher dimensions
\cite{Kaszlikowski01a,Collins01a} as well as quantum information
processing with improved security for Quantum key Distribution
\cite{Brub01a, Cerf01a} and the need to characterize these larger
quantum states. The second point provides a much larger
cross-section of the physics community with the possibility of
performing QST.

\section{1 Qubit}
To start with we will first introduce the Pauli operators using the group theoretical
definition of them as generators.  This is not crucial, though facilitates the
procedure of going to higher dimensions with more subsystems without confusing
notation changes. Hence, we can write a complete Hermitian operator basis for the
qubit space:
\begin{eqnarray}
\begin{array}{cc}
I \equiv \hat{\lambda}_0 = \left[\begin{array}{cc} 1 & 0 \\ 0 & 1
\end{array}\right] & X \equiv \hat{\lambda}_1 =
\left[\begin{array}{cc} 0 & 1 \\ 1 & 0 \end{array}\right] \\ \\
Y \equiv \hat{\lambda}_2 = \left[\begin{array}{cc} 0 & -i \\ i & 0
\end{array}\right] & Z \equiv \hat{\lambda}_3 =
\left[\begin{array}{cc} 1 & 0 \\ 0 & -1
\end{array}\right],\label{eq:su2}
\end{array}
\end{eqnarray}
corresponding to the $2\times 2$ Identity operator, $\hat{\lambda}_0$, and the generators of the SU(2) group, $\hat{\lambda}_j$, $j$ = 1,2,3. The reason for denoting these with $\hat{\lambda}_j$ will become apparent as we go to higher dimensions.  For a single qubit we can always write the density matrix as
\begin{eqnarray}
\hat{\rho}_2 = \frac{1}{2} \sum_{j=0}^{3} r_{j} \hat{\lambda}_{j},
\;\;\;\; r_{j}\in \Re.
\label{eq:qubitDM}
\end{eqnarray}
As the generators of SU(2) are all traceless operators, the normalization of the density matrix
$\hat{\rho}_2$ requires $r_0$ set to one, leaving the other parameters $r_{j=1,\ldots, 3}$
constrained only by $r_1^2+r_2^2+r_3^2\leq 1$.  The terms, $r_{j}$, can be determined from the
expectation value of the operators such that
$r_{j} = \langle \hat{\lambda}_{j}\rangle = {\rm Tr}[\hat{\rho}_{2}\hat{\lambda}_{j}] $.
Thus the single qubit density matrix has the form,
\begin{eqnarray}
\hat{\rho}_2 = \frac{1}{2}\left[\begin{array}{cc}1 +
\langle\hat{\lambda}_3\rangle & \langle\hat{\lambda}_1\rangle -
i\langle\hat{\lambda}_2\rangle \\ \langle\hat{\lambda}_1\rangle +
i\langle\hat{\lambda}_2\rangle & 1 -
\langle\hat{\lambda}_3\rangle\end{array}\right]\label{eq:qubitDM1}
\end{eqnarray}
Theoretically, only three measurements are required to define the
qubit density matrix. The fourth measurement, $\hat{\lambda}_0$,
is practically necessary, as it allows renormalization of the
count statistics to compensate for various experimental biases.
The experimental data and the calculation of the expectation
values $\langle \hat{\lambda}_{j}\rangle$ may lead to negative
eigenvalues for the density matrix even though ${\rm Tr}
\left[\hat{\rho}_2\right]=1$. This is due to the intrinsic
uncertainty in experiments, however the mathematical expression
(\ref{eq:qubitDM}) allows such nonphysical states (without the
constraint $r_1^2+r_2^2+r_3^2\leq 1$ ).  By using a maximally
likelihood technique \cite{James01a}, a physical density matrix
can be derived.

\begin{figure}[!htb]
\center{\epsfig{figure=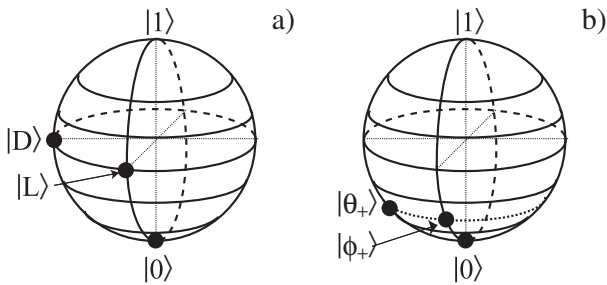,width=80mm}} \caption{Schematic of
measurements on the characteristic sphere (e.g. Poincar\'e or
Bloch ) for qubit quantum state tomography In a), an orthogonal
set of $|0\rangle, |D\rangle$ and $|L\rangle$ is shown while in b)
a non-orthogonal set $|0\rangle, |\theta_+\rangle$ and
$|\phi_+\rangle$ is shown.} \label{fig1}
\end{figure}

We note that though the SU(2) generators described above do not correspond to any
physical state, we can always write these operators in conjunction with the
Identity $\hat{\lambda}_{0}$ as a linear combination of physical basis state
density operators.  In spin systems this Pauli group provides a perfectly reasonable
set of observables, however in optics this is not the case.  In optics a more common
example could be the polarization basis,
\begin{eqnarray}\begin{array}{cc}
|H\rangle\langle H| = \frac{1}{2}[\hat{\lambda}_{0}
+\hat{\lambda}_{3}]& |V\rangle\langle V| =
\frac{1}{2}[\hat{\lambda}_{0} -\hat{\lambda}_{3}] \\
|D\rangle\langle D| = \frac{1}{2}[\hat{\lambda}_{0}+
\hat{\lambda}_{1}] & |L\rangle\langle L| =
\frac{1}{2}[\hat{\lambda}_{0} -\hat{\lambda}_{2}] \end{array},
\end{eqnarray}
where, in the computational basis,
$|H\rangle=|0\rangle$, $|V\rangle=|1\rangle$,
$|D\rangle = [|0\rangle + |1\rangle]/\sqrt{2}$ and
$|L\rangle = [|0\rangle + i|1\rangle]/\sqrt{2}$.  The three orthogonal measurements
are $|H\rangle, |D\rangle$ and $|L\rangle$ (depicted in Fig (\ref{fig1})).\\
Regardless of what orthogonal measurements we choose, we can always write
%\begin{eqnarray}
$\hat{\lambda}_{j} = \sum_{k}a_{jk}\hat{\rho}_{k}$
%\end{eqnarray}
for some other set of operators $\hat{\rho}_{k}$.  State tomography may then be
performed by measuring the expectation values
$a_{jk} = \langle \hat{\rho}_{k}\rangle= {\rm Tr}[\hat{\rho}_2\hat{\rho}_{k}]$.

\subsection{Non orthogonal state tomography}

In the state tomography that has been previously discussed we had assumed that we could
measure observables at orthogonal points on the characteristic sphere.
(For instance $|H\rangle, |D\rangle, |L\rangle$ in Fig (\ref{fig1}).).
In many practical situations the method of achieving these measurements
could be a single qubit rotation followed by a measurement on $|0\rangle$,
more explicitly, single qubit rotation would be necessary from
$|0\rangle + |1\rangle$ and $|0\rangle + i|1\rangle$ to $|0\rangle$.
One could envisage many practical situations where it is difficult
to perform these large single qubit rotations to the $|0\rangle$ state.
Does this mean that state tomography can not be performed? The answer is no,
state tomography can also be performed if one has access only to a small solid
angle on the characteristic sphere.  For ideal measurements, one still needs to
make a set of three measurements that project onto $|0\rangle$ and
\begin{eqnarray}
|\theta_+\rangle=\frac{1}{\sqrt{2}}[\cos\theta |0\rangle &+ &
\sin\theta|1\rangle]\label{eq:nonorth12} \\
|\phi_+ \rangle =\frac{1}{\sqrt{2}}[\cos\phi |0\rangle &+ & i
\sin\phi|1\rangle]\label{eq:nonorth34},
\end{eqnarray}
where $\theta,\phi$ can be small.  Thus we only require a small
perturbation about some accessible point on the characteristic
sphere (see Fig(\ref{fig1}b). This observation is likely to be
important in experiments where qubit rotation is more demanding
than measurement in the logical basis, such as flux qubit systems.

Naturally, as the measurement axes tend further away from
orthogonal, the uncertainties for a fixed number of measurements
will grow accordingly, or alternatively, achieving a target
uncertainty in the state reconstruction will require a larger
number of measurements.

Consider arbitrary states, $|\psi_{\nu}\rangle$, such that a
projection measurement is represented by $\hat{\lambda_{\nu}} =
|\psi_{\nu}\rangle \langle \psi_{\nu}|$. The count statistics
arise from a series of these measurements. Correspondingly the
average counts from a series of measurements will be
\begin{eqnarray}
     n_{\nu} = {\cal N}\langle \psi_{\nu}|\hat{\rho}|\psi_{\nu}
     \rangle
\end{eqnarray}
where ${\cal N}$ is a constant that will be dependent on
experimental factors such as detection efficiencies. The measured
counts, $n_{\nu}$, are statistically independent Poissonian random
variables and hence we assume that they will satisfy
\begin{eqnarray}
\overline{\delta n_{\nu}\delta n_{\mu}} = n_{\nu}\delta_{\nu\mu}.
\end{eqnarray}
This now allows us to consider how these statistics will vary with
respect to the nonorthogonal measurements.

The difference in count statistics when measuring with orthogonal
states and when using nonorthogonal states will be proportional to
the overlap of the two states \cite{Dieks88a}.  We now denote the
measurement statistics resulting from projecting onto one of a set
of nonorthogonal states, $|\psi_{\nu}'\rangle$, as $n_{\nu}'$.
Hence we find that the counts for nonorthogonal measurements are
related to the orthogonal in the following manner,
\begin{eqnarray}
n_{\nu}' = {\cal N}|\langle
\psi_{\nu}|\psi_{\nu}'\rangle|^{2}\langle
\psi_{\nu}|\hat{\rho}|\psi_{\nu} \rangle = n_{\nu}|\langle
\psi_{\nu}|\psi_{\nu}'\rangle|^{2}
\end{eqnarray}
with the errors appropriately scaled and given by
\begin{eqnarray}
\overline{\delta n_{\nu}'\delta n_{\mu}'} =
\frac{n_{\nu}\delta_{\nu\mu}}{|\langle
\psi_{\nu}|\psi_{\nu}'\rangle|^{2}}.
\end{eqnarray}
The counts and the errors all revert to the orthogonal case as
$|\langle \psi_{\nu}|\psi_{\nu}'\rangle|^{2} \rightarrow 1$.

\section{Generalization to qudits}

We introduced the qubit tomography in terms of the SU(2) generators. Let us now
consider a state with $d$ levels.  Firstly we prepare the generators for SU($d$)
systems and thereby construct the density matrices for a qudit system.  For
convenience we use the su algebra but we will denote the algebra for a
$d$-dimensional system as su($d$). The generators of SU($d$) group may be conveniently
constructed by the elementary matrices of $d$-dimension,
$\{ e^k_j| k,j =1\ldots d\}$.  The elementary matrices are given by
\begin{eqnarray}
     (e^{k}_{j})_{\mu \nu} = \delta_{\nu j}\delta_{\mu k},
     \hspace{1cm} 1 \le \nu,\mu \le d, \label{eq:elem}
\end{eqnarray}
which are matrices with one matrix element equal to unity and all others equal to
zero.  These matrices satisfy the commutation relation:
\begin{equation}
[e^i_j,e^k_l]=\delta_{k,j}e^i_l-\delta_{il}e^k_j.
\end{equation}
There are $d(d-1)$ traceless matrices,
\begin{eqnarray}
\Theta _{j}^{k} &=& e^{k}_{j} + e^{j}_{k} \\
\beta _{j}^{k} &=& - i(e^{k}_{j} - e^{j}_{k}), \;\;\; 1\le k < j \le
d\label{eq:genoff}
\end{eqnarray}
which are the off-diagonal generators of the SU($d$) group.  We add the $d-1$
traceless matrices
\begin{eqnarray}
     \eta_{r}^{r} = \sqrt{\frac{2}{r(r+1)}}\left[\sum_{j=1}^{r}
     e^{j}_{j} - re^{r+1}_{r+1} \right]\label{eq:genon}
\end{eqnarray}
as the diagonal generators and obtain a total of $d^2-1$ generators. SU(2) generators
are, for instance, given as
$\{ X= \Theta _{2}^{1} = e^{1}_{2} + e^{2}_{1},
Y = \beta _{2}^{1} = -i(e^{1}_{2} - e^{2}_{1}),
Z = \eta_{1}^{1} = e^{1}_{1} - e^{2}_{2}\}$.

We now define the $\lambda$-matrices, this is how we labelled the Pauli matrices in
Eq.(\ref{eq:su2}),
\begin{eqnarray}
     \lambda_{(j-1)^{2}+2(k-1)} &=& \Theta _{j}^{k} \label{eq:lambdadef1}\\
     \lambda_{(j-1)^{2}+2k-1} &=& \beta _{j}^{k} \label{eq:lambdadef2}\\
     \lambda_{j^{2}-1} &=& \eta_{j-1}^{j-1} \label{eq:lambdadef3},
\end{eqnarray}
which, as shown previously, produce the $X, Y, Z$ operators of the
SU(2) group and so on for higher dimensions. In conjunction with a
scaled  $d$-dimensional identity operator these form a complete
hermitian operator basis.

It is then straightforward from Eq.(\ref{eq:qubitDM}) to see that
a density matrix $\rho_d$ can be a linear combination of the
generators as
\begin{equation}
\hat{\rho}_d = \frac{1}{d} \sum_{j=0}^{d^2-1} r_{j} \hat{\lambda}_{j}.
\end{equation}
This $\rho_d$ is a density matrix of dimension $d$, a qudit, and the coefficient $r_0$ is
one for the normalization.  The condition
${\rm Tr} \left[\rho_d^2\right]\leq 1$ requires
$\sum_{j=1}^{d^2-1} r_{j}^2 \leq d (d-1)/2$.

Now let us extend these results to $n$-qudits.  It was shown that
for multiple qubits we only had to consider a space of operators
defined by the tensor product of the generators, SU(2) $\otimes$
SU(2) $\otimes \ldots \otimes$ SU(2) where we have included
$\lambda_0$ (the normalized identity matrix) with the normal SU(2)
generators\cite{James01a}. For two qudits, a density matrix
$\rho_{2\; d}$, which has dimension $d^2$, can be expanded
similarly. All combinations of the tensor products of the
$\lambda$-matrices (complimented with $\lambda_0$),
$\lambda_{j1}\otimes \lambda_{j2}$, are linearly independent to
each other. Hence, the expression of the density matrix $\rho_{2\;
d}$ may be written in terms of $\lambda$-matrices,
\begin{equation}
\hat{\rho}_{2\; d} = \frac{1}{d^{2}} \sum_{j1,j2 =0}^{d^2 -1}
r_{j1,j2} \hat{\lambda}_{j1}\otimes \hat{\lambda}_{j2}.
\end{equation}
Similarly this expression can be generalized to density matrices of $n$ qudits,
that is
\begin{eqnarray}
\hat{\rho}_{nd} = \frac{1}{d^{n}} \sum_{j1\ldots jn =0}^{d^2 -1}
r_{j1\ldots jn} \hat{\lambda}_{j1}\otimes \ldots
\otimes\hat{\lambda}_{jn}.
\label{eq:nquditDM}
\end{eqnarray}
\begin{figure}[!thb]
\center{\epsfig{figure=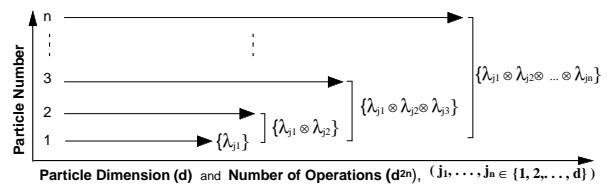,width=80mm}} \caption{The
measurement scaling for tomography on n-qudits results from the
necessity to measure every basis state on every subsystem in every
permutation. The measurements scale as $d^{2n}-1$ where $d$ is the
{\it particle dimension\/}, e.g. $d=2$ for a qubit, and $n$ is the
number of particles.} \label{fig:complex}
\end{figure}
The tomography on such a state is only restricted by the patience of the
experimentalist to determine the expectation values for the systems observables,
\begin{eqnarray}
     r_{j1\ldots jn} = \langle \hat{\lambda}_{j1}\otimes \ldots
     \otimes \hat{\lambda}_{jn} \rangle.
\end{eqnarray}
There will we $d^{2n}-1$ measurements required if we assume
perfect detection. Fig.(\ref{fig:complex}) illustrates the scaling
catastrophe that occurs for multiple parties of higher-dimensional
states. The key concept in both the extension to higher
dimensional states and to more subsystems is that for each
subsystem we need to measure every basis state on every subsystem
in every permutation.

However if some structure is known about the state, then the
number of measurements can be reduced. For example if we are
confident that we are only ever dealing with a pure state then the
number of measurements is significantly reduced and the scaling of
measurements more so. QST for two qubits normally requires 15
measurements. If we know this state is pure this is reduced to 6:
3 on the diagonal; and 3 on the anti-diagonal.  (In the case where
we know the state to be, say, one of the Bell states, then this is
reduced further to just 2).  So in general for pure states we only
require $2(d^n - 1)$ measurements to reconstruct the density
matrix.

The principle of nonorthogonal state tomography carries through to
the higher dimensional cases in exactly the same way that it does
for normal tomography using orthogonal states as do the
considerations with respect to errors. Also a detailed discussion
regarding the sources of error and there effect was outlined by
James {\it et al.} \cite{James01a} which was derived for the
qubits but is equally valid for qudits by simple substitution and
appropriate change in the summation ranges.

\section{Qutrits}

As a specific example of how we can implement higher dimension tomography consider a
qutrit, $d = 3$ dimensional, state.  We can write this as
\begin{eqnarray} \hat{\rho}_3 = \frac{1}{3}
\sum_{j=0}^{8} r_{j} \hat{\lambda}_{j}\label{eq:qutritDM}
\end{eqnarray}
where the $\hat{\lambda}_{j}$ are now the SU(3) generators and an Identity
operator $\hat{\lambda}_0$. For SU(3) the set of generators are
\begin{eqnarray}
\begin{array}{cc}
\hat{\lambda}_1 = \left[\begin{array}{ccc} 0&1&0\\1&0&0\\0&0&0
\end{array} \right] & \hat{\lambda}_2 = \left[\begin{array}{ccc}
0&-i&0\\i&0&0\\0&0&0 \end{array} \right]\\ \\
\hat{\lambda}_3 = \left[\begin{array}{ccc} 1&0&0\\0&-1&0\\0&0&0
\end{array} \right] & \hat{\lambda}_4 = \left[\begin{array}{ccc}
0&0&1\\0&0&0\\1&0&0
\end{array}
\right]\\ \\
\hat{\lambda}_5 = \left[\begin{array}{ccc} 0&0&-i\\0&0&0\\i&0&0
\end{array}
\right] & \hat{\lambda}_6 = \left[\begin{array}{ccc}
0&0&0\\0&0&1\\0&1&0
\end{array}
\right]
\label{eq:qutritgen} \\ \\
\hat{\lambda}_7 = \left[\begin{array}{ccc} 0&0&0\\0&0&-i\\0&i&0
\end{array}\right] & \hat{\lambda}_8 =
\frac{1}{\sqrt{3}}\left[\begin{array}{ccc} 1&0&0\\0&1&0\\0&0&-2
\end{array}\right]
\end{array}
\end{eqnarray}
which have been determined using the definitions of
Eq.(\ref{eq:lambdadef1}-\ref{eq:lambdadef3}) and the corresponding elementary
matrices of Eq.(\ref{eq:elem}).

Once we have the expectation values for these operators then the
density matrix can be reconstructed in the same way that it was
done for the qubit in Eq.(\ref{eq:qubitDM1}):
\begin{widetext}
\begin{eqnarray}
\hat{\rho}_{3} = \frac{1}{3}\left[\begin{array}{ccc}
1+\frac{\sqrt{3}}{2}(\langle\hat{\lambda}_8\rangle +
\sqrt{3}\langle\hat{\lambda}_3\rangle) &
\frac{3}{2}(\langle\hat{\lambda}_1\rangle -
i\langle\hat{\lambda}_2\rangle) &
\frac{3}{2}(\langle\hat{\lambda}_4\rangle -
i\langle\hat{\lambda}_5\rangle) \\
\frac{3}{2}(\langle\hat{\lambda}_1\rangle +
i\langle\hat{\lambda}_2\rangle) &
1+\frac{\sqrt{3}}{2}(\langle\hat{\lambda}_8\rangle -
\sqrt{3}\langle\hat{\lambda}_3\rangle) &
\frac{3}{2}(\langle\hat{\lambda}_6\rangle -
i\langle\hat{\lambda}_7\rangle)\\
\frac{3}{2}(\langle\hat{\lambda}_4\rangle +
i\langle\hat{\lambda}_5\rangle) &
\frac{3}{2}(\langle\hat{\lambda}_6\rangle +
i\langle\hat{\lambda}_7\rangle) & 1
-\sqrt{3}\langle\hat{\lambda}_8\rangle \end{array} \right]
\label{eq:qutritDMexplicit}
\end{eqnarray}
\end{widetext}
The most direct way to do this is to measure the expectation
values for the $\hat{\lambda}$-operators. However if this is not
possible let us assume we can measure some set of basis states.
Consider an arbitrary, but complete, set of basis states
$\{|\psi_{i}\rangle \}$ with the associated projection operators
$\{\hat{\mu}_{i} = |\psi_{i}\rangle \langle \psi_{i}| \}$.  These
can be linearly related, via a $d^2\times d^2$ matrix $A$, to the
$\lambda$-matrices, $\hat{\mu}_{i} =
\sum_{j}A_i^{j}\hat{\lambda}_{j}$.  We can thus consider
measurement outcomes,
\begin{eqnarray}
n_{i} = {\cal N}\langle \psi_{i}|\hat{\rho}|\psi_{i}\rangle&=& {\cal
N}{\rm Tr}[\hat{\rho}\;\hat{\mu}_i ] \nonumber \\
 &=& {\cal N}\sum_{j=0}^{8}A_{i}^{j}{\rm Tr}[\hat{\rho}\;\hat{\lambda}_j] \nonumber \\
& = & {\cal N}\sum_{j=0}^{8}A_{i}^{j} r_{j}\label{eq:counts},
\end{eqnarray}
where ${\cal N}$ is again a constant that will be dependent on
experimental factors such as detection efficiencies. So we find,
$r_{j} = {\cal N}^{-1}\sum_{i=0}^{8}(A_{i}^{j})^{-1} n_{i}$, and
finally,
\begin{eqnarray}
\hat{\rho}_{3} = {\cal N}^{-1}\sum_{i,j=0}^8 (A_{i}^{j})^{-1}n_{i}
\hat{\lambda}_{j}\label{eq:recon3}
\end{eqnarray}
In this way the state is reconstructed from the measurement outcomes in some
arbitrary basis and the $A$-matrix which relates the measurement basis to the
$\lambda$-matrices.  This $A$-matrix will be invertible if a complete set of
tomographic measurements are made, ie. if we measure in a complete basis. The
$A $-matrix becomes the identity in the case where we use the generators.

\begin{figure}[!thb]
\center{\epsfig{figure=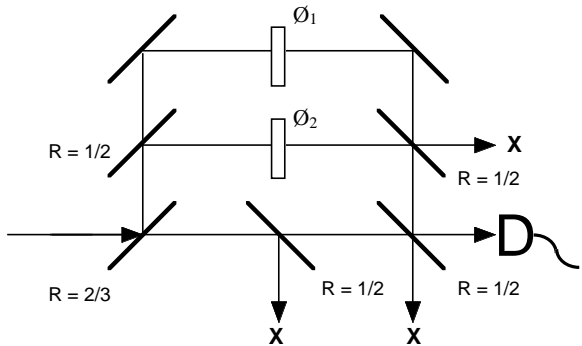,width=80mm}} \caption{A linear
optic implementation of a qutrit where the levels of the system
are encoded in the pathlength travelled. The reflectivities for
the beamsplitters, R, are given and the extra beamsplitter in the
short arm effectively balances the superposition of the output.
Phase elements in two of the arms provide the ability to consider
any balanced superposition state in the qutrit space.}
\label{fig:loqutrit}
\end{figure}

Take a physical realization of a qutrit in an linear optics regime.
Fig.(\ref{fig:loqutrit}) shows one way in which a qutrit may be realized
\cite{KwiatJMO}. The modes correspond to a photon taking the short medium or
long paths of the interferometer. The values of the reflectivities of the
beamsplitters are such that an even superposition state is generated. By varying
the phases $\phi_1$ and $\phi_2$ a complete basis can be generated,
\begin{eqnarray}
\begin{array}{cc}
\hspace{-2.25cm}|0\rangle &   \hspace{-1cm}    |1\rangle \\
\hspace{-.9cm}|0\rangle + ¦1\rangle + ¦2\rangle & \\
 |0\rangle +
\alpha|1\rangle+\alpha^2|2\rangle & \hspace{1.2cm}
|0\rangle + \alpha^2|1\rangle +\alpha|2\rangle \\
\hspace{-.4cm}|0\rangle + |1\rangle + \alpha|2\rangle &
\hspace{1cm} |0\rangle +\alpha|1\rangle+ |2\rangle \\
\hspace{-.3cm}|0\rangle + |1\rangle + \alpha^2|2\rangle &
\hspace{1.1cm}|0\rangle + \alpha^2|1\rangle +|2\rangle,
\end{array}
\end{eqnarray}
where $\alpha = e^{2\pi i/3}$.
We can then utilize another 3-arm interferometer as that shown in
Fig.(\ref{fig:loqutrit}) to rotate and perform projective
measurements on the qutrit. Therefore one can perform a series of
these projective measurements and, via the procedure outlined in
Eq.(\ref{eq:counts}-\ref{eq:recon3}), reconstruct the qutrit.

The same procedure applies regardless of the architecture provided
we measure a complete set of states. To take another optical
example, orbital angular momentum could be used to realize qutrits
(and indeed, qudits), with holographic plates generating the
qutrit superpositions and holographic interferometers acting as
analyzers.

If we now further extend this to two qutrits, which may be entangled,
\begin{eqnarray}
\hat{\rho}_{23} = \frac{1}{9}\sum_{j,k} r_{jk} \hat{\lambda}_{j}
\otimes \hat{\lambda}_{k},
\end{eqnarray}
we can consider operators of the form $ \hat{\lambda}_{j} \otimes \hat{\lambda}_{k}$,
or linearly related operators,
\begin{eqnarray}
\hat{\mu}_i \otimes \hat{\mu}_j = \sum_{k,l=0}^{8}
A_{ij}^{kl}\hat{\lambda}_{k}\otimes \hat{\lambda}_{l},
\end{eqnarray}
where the $i, j$ label the rows and $k, l $ the columns of the
$A$-matrix. There will now be $d^{2n}-1 = 80$ measurements to be
made. Therefore, as we did for one qutrit, we can again consider
the measurement outcomes for states of the form
$\{|\psi_{i}\rangle \otimes |\psi_{j}\rangle = |\psi_{ij}
\rangle\}$ with the associated projection operators
$\{\hat{\mu}_{ij} = \hat{\mu}_{i}\otimes\hat{\mu}_{j} =
|\psi_{ij}\rangle \langle\psi_{ij}| \}$.
\begin{eqnarray}
n_{ij} = {\cal N}\langle \psi_{ij}|\hat{\rho}|\psi_{ij}\rangle&=&{\cal N}{\rm Tr}[\hat{\rho}(\hat{\mu}_i \otimes \hat{\mu}_j)]
\nonumber \\
 &=& {\cal N}\sum_{k,l=0}^{8}A_{ij}^{kl}{\rm
 Tr}[\hat{\rho}\hat{\lambda}_k \otimes \hat{\lambda}_l] \nonumber \\
& = & {\cal N}\sum_{k,l=0}^{8}A_{ij}^{kl} r_{kl}.
\end{eqnarray}
So we find, $ r_{kl} = {\cal N}^{-1}\sum_{i,j=0}^{8}(A_{ij}^{kl})^{-1} n_{ij}$, and
finally,
\begin{eqnarray}
\hat{\rho}_{23} = {\cal N}^{-1}\sum_{i,j,k,l=0}
(A_{ij}^{kl})^{-1}n_{ij} \hat{\lambda}_{k} \otimes \hat{\lambda}_{l}
\end{eqnarray}
\begin{figure}[!thb]
\center{\epsfig{figure=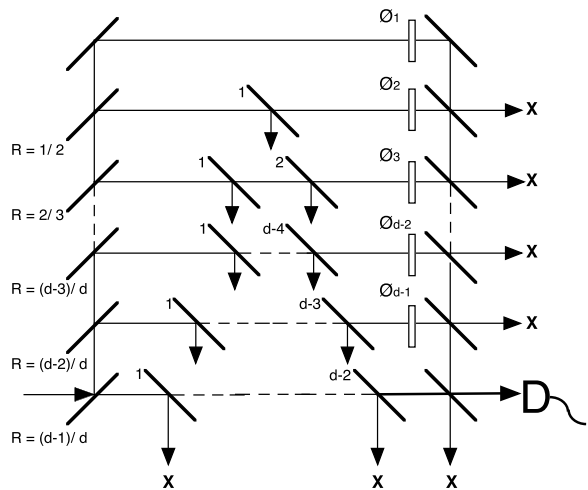,width=80mm}} \caption{A state
generation, or measurement, scheme for qudits using linear optical
elements. The beamsplitter reflectivities and phases are in
complete analogy to the description in Fig.(\ref{fig:loqutrit})
for the qutrit generation. For generation and measurement the
number of elements scale as  $d^2+ 3d$.} \label{loscale}
\end{figure}
We can then reconstruct the density matrix for the state using the experimental
measurement outcomes, $n_{i,j}$, and this $A$ matrix. Once we have the density
matrix for the entangled qutrit state we can then consider questions of purity
and entanglement. We refer the reader to \cite{Caves00a} which gives a thorough
exposition with respect to characterizing entangled qutrits that is of relevance
to both pure and mixed states.

This change of basis is completely general and allows us to consider the
reconstruction of any discrete system.  We can now use: the generators; any
orthonormal physical basis set; or, more importantly, in the case where we have
limited access to the state space, a non-orthogonal basis.

As mentioned previously there is significant motivation to study entangled
$d$-dimensional states and with the reconstruction of the complete density
matrix many important state characteristics can be determined. In practice however,
the dimensions will be restricted due to the complexity in implementing the
measurements of the $d$-dimensional state.

In the case of generating a qudits using a linear optical elements
the number of elements required to generate and hence also measure
these higher dimensional states increases rapidly.
Fig.(\ref{loscale}) shows the general scaling for a system to
generate qudits in a linear optics regime. For this implementation
the state generation and measurement requires $d^2+ 3d$ elements
for each qudit. The probability of producing these state scales as
$(1/2)^{d-1}$ and similarly for its measurement. Similar
complexity issues will be relevant regardless of the architecture.

\section{Conclusion}
We have given a simple yet illustrative account of  Quantum State
Tomography for discrete systems, from a single qubit with an
orthonormal measurement basis to multipartite-multidimensional
systems with limited access to measurements in the Hilbert space. The specific
example for the qutrit highlights the similarities and differences
in going to higher dimensions whilst constructing an intuitive
framework for the Quantum Information experimentalist to work.
Primarily it is hoped that we have made QST relevant and
accessible  to a wider cross-section of the physics community. QST
can provide a powerful tool for the experimentalist in QIS
regardless of physical implementation, be it  ion trap, quantum
dot, flux qubit, or photon, to name but a few.

\section*{Acknowledgments}
RTT would like to thank J. Altepeter for fruitful discussions and
acknowledge the hospitality of Hewlett Packard.

\end{document}